%
%
\voffset=.4in
\magnification=\magstep1
\parskip 3pt plus 1pt minus 1pt
\centerline{[ Modern Physics Letters A, Vol.\thinspace {\bf 13}, 
1265-1280 (1998) ]}
\vskip 3pc
\centerline {\bf A MODEL OF NON-LOCAL QUANTUM ELECTRODYNAMICS :}
\centerline {\bf TIME'S ARROW AND EPR-LIKE QUANTUM CORRELATION}
\vskip 1pc
\centerline {T K Rai Dastidar\footnote *{Department of Materials 
Science. E-mail : mstkrd@mahendra.iacs.res.in}~~and~~Krishna Rai 
Dastidar\footnote 
{**}{Department of Spectroscopy. E-mail : spkrd@mahendra.iacs.res.in}}
\centerline {\it Indian Association for the Cultivation of Science,
Calcutta 700032, India}
\vskip 2pc
\centerline {\bf ABSTRACT}
\vskip 1pc
A recent experiment with squeezed light has shown that two-photon absorption
by an atom can occur with a {\it linear intensity dependence\/}. We point
out that this result verifies a prediction made by us a decade back from an
analysis of a non-local model of Quantum Electrodynamics. This model had
earlier been proposed by us in an {\it ad hoc\/} manner to interpret 
certain features of multiphoton double 
ionisation and above-threshold ionisation in an atom placed in a strong 
laser field ; in this paper we show that the model can be obtained 
field-theoretically by demanding covariance of the field Lagrangian
under a non-local U(1) gauge transformation. We also show that the
model makes direct contact 
with squeezed light, and thus allows us to describe these two completely
different scenarios from a unified point of view. We obtain a fundamentally 
new result from our non-local model of QED, namely that {\it only the past, 
but not the future, can influence the present\/} --- thus establishing a 
non-thermodynamic arrow of time. We also show that correlations within a 
quantum system {\it should necessarily be of the Einstein-Podolsky-Rosen
{\rm (EPR)}-type\/}, a result 
that agrees with Bell's theorem. These results follow from the simple 
requirement of energy conservation in matter-radiation interaction. 
Furthermore, we also predict new and experimentally verifiable results on 
the basis of our model QED.
\vfill\eject
\centerline {\bf 1.\quad INTRODUCTION}
\vskip 1pc
In a beautiful experiment, Georgiades et al [1] generated a squeezed vacuum 
field by an optical parametric amplifier and focussed it onto a cloud of
cesium atoms in a magneto-optic trap. They found that the rate of the 
two-photon transition $6S_{1/2}\to 6D_{5/2}$ deviated from the expected 
quadratic intensity dependence to a form $R=\alpha I+\beta I^2$, where $I$
was the squeezed vacuum intensity ; the linear term dominated at low 
intensities, and the quadratic term at higher intensities. Our conventional
wisdom tells us that this feature of linear intensity dependence of 
two-photon transitions is shared neither by thermal light nor by coherent 
light.

In this paper we draw the reader's attention to a non-local model of
Quantum Electrodynamics (QED), suggested quite a few years ago by us [2] 
in which a possibility of this linear intensity dependence of two-photon
transitions was pointed out for the first time, albeit in a different 
context of multiphoton transitions and above-threshold ionisations in an
atom placed in a laser field. We were led to this model
of non-local QED through an analysis of the behaviour of an atom 
placed in an intense laser field under certain specified conditions ---
namely, that two distinct photon absorption events by the atomic electron/s
be {\it correlated in phase\/} --- and we obtained this two-photon 
transition with a linear intensity dependence as the central result of 
the paper. In [2], however, the non-local potential was introduced in an
{\it ad hoc\/} manner ; in the present paper we provide a field-theoretical 
basis for the same and briefly repeat the analysis made in [2]. We
bring out certain parallelisms between the two contexts (intense laser 
field and squeezed vacuum field) and show that the aforesaid model of
non-local QED can be applied to both. As a particular result of this
application, we find 
that the time evolution of an (atom + field) system under the said 
conditions shows a remarkable resemblance to a {\it squeezed vacuum\/}. 
(As we shall show, apart from pre-multiplying factors, they turn 
out to be {\it identical\/}). We further show that a specific result for 
a single-electron atom placed in a non-local electromagnetic field 
exactly matches the abovementioned experimental result of two-photon 
transition in Cesium placed in a squeezed vacuum.  

Subsequent to our work [2], a few
studies on squeezed light (e.g. Gea-Banacloche [3], Javanainen and Gould [4])
predicted this linear intensity dependence from certain special models of
the four-field correlation function in squeezed light fields.
Our approach needs no such squeezed-field-specific
modelling, but is valid for all (classical as well as non-classical)
radiation fields where phase-correlated photon absorption/emission 
can take place in interaction with matter, as we shall show below.

Consider an intense classical (thermal or laser) light incident on an 
atom, and let a pair of photon absorption events occur at two different 
space-time points $({\bf r},t)$ and $({\bf r'},t')$, the interval between 
which is very close to zero, i.e.
$\vert{\bf r}-{\bf r}'\vert \sim c\vert t-t'\vert$. This situation
can arise when, say, the outer electrons of a Rydberg atom absorb a pair
of photons within a time $\delta t \ll {1\over\omega}$. For laser intensities
of order $10^{10}$ ${\rm W/cm^2}$ or higher, this is commonplace, if we 
remember that in the visible range, a flux of one photon $a_0^{-2} t_0^{-1}$
corresponds to $\sim$ 5 $\times$ $10^{14}$ ${\rm W/cm^2}$ of intensity
($a_0$ and $t_0$ being the atomic units of length and of time respectively).
We have here a situation where two events, which may have occurred far 
apart in space, are correlated in phase ; indeed, if the events are 
spatially separated, the correlation assumes the character of an EPR-like 
``quantum entanglement''.

Consider now a different scenario : radiation from an optical
parametric oscillator generating a phase-squeezed vacuum field is absorbed 
by a three-level atom having a principal quantum number $\ge$ 5. If the
squeezed vacuum is produced by parametric frequency down-conversion, where
an incident pump photon has split up into two lower-frequency photons
(signal and idler) which constitute a highly correlated photon pair,
a two-photon absorption similar as observed in [1] occurs ; i.e.
two photons in exact (or nearly exact) phase correlation can be absorbed
in two events widely separated in space even at low intensities. 

The feature common to these two entirely different scenarios is evident.
The latter scenario is directly linked to the abovementioned experimental 
finding in [1], while the former is linked to 
phase-correlated multiphoton excitation/ionization
{\it in a regime where, however, perturbation theory holds\/}. In this
paper we show that our model in [2] can be used, by exploiting this common
feature of correlated atomic response to radiation at two points far apart 
in space, to give a theoretical description of these two different scenarios 
from a unified viewpoint. As already mentioned, the linear intensity 
dependence of two-photon processes comes about as a natural result of this 
model, and we shall show how this linear intensity dependence serves to
explain several features long known in multiphoton double ionisation
as well as in above-threshold ionisation, and also serves to predict
some new results. 
\vskip 1pc
\centerline {\bf 2.\quad THEORY}
\vskip 6pt
In the usual simple harmonic oscillator mode expansion of the electromagnetic
field potential {\bf A}, an integral number of photons (0,1,2,...)
is associated with each mode. All these modes are orthogonal and
independent of each other, and  since photons obey the Bose statistics, each 
mode can be occupied by any arbitrary number of photons. However, in situations
such as described above where a {\it phase-correlated behaviour\/} of 
photons becomes manifest, one might think whether a different, 
{\it correlated\/} type of mode expansion could better describe
the situation, much as a correlated wave-function gives a better 
representation of, say, the Helium atom than the orbital product
wavefunction does. 
Since there is no force acting between photons, the observed ``correlation''
between them is not due to any ``configuration interaction'' as in the case 
of a multi-electron system, but due either to the presence of certain 
non-classical features in the radiation field introduced by a non-linear
medium (e.g. squeezing), or to correlated interaction of the radiation 
field with the atomic electrons, or both.

We start from the basic fact that in both these scenarios the radiation
field (either as generated, or as detected by the atomic electrons)
enjoys a large second-order coherence, i.e. two photons are {\it coherently\/}
absorbed by the atomic electrons within a vanishingly small phase difference
\thinspace $\Delta\phi=\omega\delta t$.  
In a study of two-photon transition using a
phase-squeezed light within the framework of the usual QED, Javanainen 
and Gould [4] observed that the 
``four-field correlation function $G^{(2)}({\bf r},t;{\bf r}',t') =
\langle E^-({\bf r},t)E^-({\bf r}',t')E^+({\bf r}',t') 
E^+({\bf r},t)\rangle$, essentially the joint probability that two detectors 
placed at {\bf r} and ${\bf r}'$ will record photon 
counts at times $t$ and $t'$, does not always suffice to predict the
two-photon transition rate. In fact, we need a more general correlation
function\dots We resort to heuristic modelling of the field
correlations''.

In both [3] and [4] the authors gave new four-field correlation functions
derived from specific properties of squeezed light. 
In the present paper we show that neither the use of such specific properties 
nor any heuristic modelling is necessary, if we adopt the earlier proposal
of attempting a correlated description of the electromagnetic field. 

We show below that the non-local model of QED as expounded in [2] serves
to bring out the feature of (phase) correlation between two (or more)
photon modes. Consider two zero-order dressed states 
$\vert\Psi_f\rangle = \vert\psi_f\rangle\vert n_f\rangle$ and 
$\vert\Psi_i\rangle=\vert\psi_i\rangle\vert n_i\rangle$.
In the Coulomb gauge the radiation interaction matrix element 
coupling these two states is $M_{fi}(t) = -{e\over {mc}}\langle
\psi_f({\bf r}),n_f\vert {\bf p}.{\bf A(r},t)\vert\psi_i({\bf r}),
n_i\rangle$. As in our earlier work [2], we introduce a non-local potential 
${\cal A}({\bf r},t,{\bf r'},t')$ such that the above matrix element becomes
$$M_{fi}(t)=-{e\over {mc}}\langle\psi_f(x),n_f\vert{\bf p}.
{\cal A}(x,x')\vert\psi_i(x'),n_i
\rangle\qquad {\rm where}\quad x\equiv ({\bf r},t),\quad x'\equiv 
({\bf r}',t'),\eqno(1)$$
i.e. the interaction at the point $x$ is correlated with the interactions 
at all other points $x'$. This non-local interaction had been defined in [2]
in an {\it ad hoc\/} manner ; in this paper we show (see Appendix A) 
how such a potential can be introduced field-theoretically by demanding 
covariance under a non-local U(1) gauge transformation.

The integration in (1) runs over $d{\bf r}$ and 
$dx'$, and in this paper we set the limits 
of the time integral $dt'$ either (i) from $-\infty$ to 
$t-{{\vert {\bf r}-{\bf r}'\vert}\over c}$ (giving the retarded 
interaction), or (ii) from $t+{{\vert {\bf r}-{\bf r}'\vert}\over c}$ 
to $\infty$ (which gives the advanced interaction), leaving out EPR
``quantum entanglement''-like correlations for the present. We shall discuss
the question of causality later and shall find that this restriction
turns out to be unnecessary. 

We can equivalently use a similar non-local form of the multipolar interaction 
at the space-time point $x$ :
$$-e\int {\bf r}.{\cal E}(x,x')\,dx'\eqno(1')$$
Although nothing can be said about the specific mathematical forms of these
non-local potentials/field strengths, we will find that we can go a long
way to predict specific consequences of this general non-local field.
In order to quantise this non-local field, we proceed by carrying out a 
Fourier expansion [5] of this non-local potential over the usual photon modes :
$${\cal A}(x,x')=\sum_{{\bf k}_1\lambda_1}\sum_{{\bf k}_2\lambda_2}\left [C_{12}
\hat\epsilon_{12}(\hat r,\hat r')\exp (i{\bf k}_1.{\bf r}+i{\bf k}_2.{\bf r}'
-i\omega_{{\bf k}_1}t-i\omega_{{\bf k}_2}t')+{\rm c.c.}\right ]\eqno(2)$$
where we have written $C_{12} \equiv C_{{\bf k}_1\lambda_1{\bf k}_2\lambda_2}$,
and $\hat\epsilon_{12} \equiv \hat\epsilon_{{\bf k}_1\lambda_1{\bf k}_2
\lambda_2}$. In view of (1), we cannot fix the direction of the polarisation 
vector $\hat\epsilon(\hat r,\hat r')$ ``by hand'' (as in standard QED) 
without loss of generality ; this is at once obvious if we
make the dipole approximation in (1) or $(1')$. Later in this paper we
shall use for $\hat\epsilon(\hat r,\hat r')$ a power series in $\hat r$ and 
$\hat r'$ with undetermined coefficients, which is the most general form
possible. However, to quantise the field using the Coulomb gauge we 
impose for the present the condition that 
$$(\hat{\bf k}_1+\hat{\bf k}_2).\hat\epsilon=0.\eqno(3)$$ 
This restriction will be removed later.

From eqn.\thinspace (A2) in Appendix A, we note that the non-local electric 
field in the Coulomb gauge is given by
$${\cal E}(x,x') = -(1/c)\left ({\partial{\cal A}\over{\partial t}} + 
{\partial{\cal A}\over{\partial t'}}\right ) = 
{i\over c}\sum_{{\bf k}_1\lambda_1}\sum_{{\bf k}_2\lambda_2}
(\omega_{{\bf k}_1}+\omega_{{\bf k}_2})\left 
[C_{12}\hat\epsilon_{12}(\hat r,\hat r')f(x,x')-{\rm c.c.}\right ]\eqno(4)$$
where $f(x,x')$ is the phase factor in (2). The expression for the non-local
magnetic field can be found in an analogous manner. 

To obtain the field energy at a point $({\bf r},t)$ we need $E^2$ and $B^2$
which we define similarly as in (1) ; thus
$$E^2({\bf r},t) = \int \int {\cal E}(x,x').{\cal E}(x,x'')\,dx'\,dx''$$
where the limits of integration over $dt'$ and $dt''$ are determined by the 
choice of interaction (retarded or advanced). However, in the present work
$r$'s are of atomic dimensions, and we simply put $t',t''\leq t$ for
the retarded interaction and $t',t''\geq t$ for the advanced interaction.

An examination of eqns. (2) and (4) shows that in $E^2$ and $B^2$, the 
crossed terms between two modes ${\bf k}_1\lambda_1$ and ${\bf k}_2\lambda_2$
vanish unless the two modes are strongly phase-correlated. Denoting the
population fractions of such correlated-pair modes and of the independent
modes by $a_{\rm II}$ and $a_{\rm I}$ respectively, use of standard
methods in Quantum Electrodynamics leads to the following expression for
the field energy in a volume $\Omega$ :  
$$W=a_{\rm I}^2W_{\rm I}+a_{\rm II}^2W_{\rm II}\eqno(5)$$
where $W_{\rm I}$ is the usual energy summed over single modes, and 
\vskip 6pt
\hfill $W_{\rm II} = {1\over 2}\sum\limits_{{\bf k}\lambda}
\left (Y_{{\bf k}\lambda}^2+4\omega_{\bf k}^2Z_{{\bf k}\lambda}^2\right )$,
\hfill $(5')$
\vskip 6pt
\noindent where\quad $Y_{{\bf k}\lambda}=-i{\sqrt{\Omega/\pi}\over {2c}}
2\omega_{\bf k}(C_{{\bf k}\lambda}-C_{{\bf k}\lambda}^*)$,\quad 
$Z_{{\bf k}\lambda}={\sqrt{\Omega/\pi}\over{2c}}(C_{{\bf k}\lambda}+
C_{{\bf k}\lambda}^*)$. Each mode subscript here actually stands for a
correlated mode pair $({\bf k}_1\lambda_1,{\bf k}_2\lambda_2)$, and
$2\omega_{\bf k}$ stands for $\omega_{{\bf k}_1}+\omega_{{\bf k}_2}$.
$Y$ and $Z$ satisfy the canonical equations of motion with $W_{\rm II}$ as 
the Hamiltonian as in standard QED, and we now quantise
this non-local field by requiring that $Y$ and $Z$ be $q$-numbers obeying the 
commutation relation
$$\left [Z_{{\bf k}\lambda}, Y_{{\bf k}'\lambda '}\right ] = 
i\hbar\delta_{{\bf kk}'}\delta_{\lambda\lambda '}.\eqno(6)$$

We thus obtain the non-local field energy in terms of a {\it new pair of 
creation and annihilation operators\/} :
\vskip 6pt
\hfill $W_{\rm II}={1\over 2}\sum\limits_{{\bf k}\lambda}\hbar\omega_{\bf k}
\left (b_{{\bf k}\lambda}b^{\dag}_{{\bf k}\lambda}+b^{\dag}_{{\bf k}\lambda}
b_{{\bf k}\lambda}\right )$
\hfill $(5'')$
$${\rm where}\quad b_{{\bf k}\lambda}, b^{\dag}_{{\bf k}\lambda} = 
{1\over{\sqrt{2\hbar\omega_{\bf k}}}}
\left (2\omega_{\bf k}Z_{{\bf k}\lambda}\pm iY_{{\bf k}\lambda}\right ).
\eqno(7)$$ 
Equation $(5'')$ is identical in form with standard QED, but the creation 
and annihilation operators obey a {\it new commutation relation\/} :
$$\left [b_{{\bf k}\lambda},b^{\dag}_{{\bf k}'\lambda '}\right ] = 2\delta_
{{\bf kk}'}\delta_{\lambda\lambda '},\quad 
\left [b_{{\bf k}\lambda},b_{{\bf k}'\lambda '}\right ]
= \left [b^{\dag}_{{\bf k}\lambda},b^{\dag}_{{\bf k}'\lambda '}\right ] 
= 0\eqno(8)$$
as may be directly verified from (7) and (6). The following, and only the
following, results of operation of these new correlated-pair mode operators 
are consistent with the commutation rules (8) :
$$b_{{\bf k}\lambda}\vert n_{{\bf k}\lambda}\rangle = \sqrt{n_{{\bf k}\lambda}}
\vert n_{{\bf k}\lambda}-2\rangle ,\quad 
b^{\dag}_{{\bf k}\lambda}\vert n_{{\bf k}\lambda}\rangle =
\sqrt{n_{{\bf k}\lambda}+2}\vert n_{{\bf k}\lambda}+2\rangle ,\quad 
b^{\dag}_{{\bf k}\lambda}b_{{\bf k}\lambda}\vert n_{{\bf k}\lambda}\rangle = 
n_{{\bf k}\lambda}\vert n_{{\bf k}\lambda}\rangle $$
This shows that $b$ and $b^{\dag}$ are
{\it two-photon annihilation/creation operators\/}, and $b^{\dag}b$ is the
usual number operator. Once again, by creation of {\it two photons\/} we mean 
the creation of {\it one photon in each member of a correlated mode pair\/},
and similarly for annihilation. We can now write down the potential and 
field strengths in terms of these new operators ; thus
$${\cal A}(x,x') = \sum_{{\bf k}\lambda}\sqrt{{2\pi\hbar c^2}\over{\Omega
\omega_{\bf k}}}
\hat\epsilon_{{\bf k}\lambda}(\hat r,\hat r')\left [b_{{\bf k}\lambda}f(x,x') +
b^{\dag}_{{\bf k}\lambda}f^*(x,x')\right ]\eqno(9a)$$
$${\cal E}(x,x') = i\sum_{{\bf k}\lambda}\sqrt{{2\pi\hbar\omega_{\bf k}\over 
\Omega}}\hat\epsilon_{{\bf k}\lambda}(\hat r,\hat r')\left [b_{{\bf k}\lambda}
f(x,x')-b^{\dag}_{{\bf k}\lambda}f^*(x,x')\right ]\eqno(9b)$$

The results obtained so far from our non-locality postulate may be summed
up as follows : quantization of a radiation field with a high degree of
second-order coherence can depart from the usual QED and can lead to photon 
modes (more exactly: correlated mode pairs) generated by two-photon creation 
and annihilation operators. Indeed, if we let a zero-order dressed state of an 
atom $\vert\psi\rangle\vert n\rangle$ evolve in the presence of the non-local
electron-radiation interaction (1) or $(1')$, we obtain a linear superposition 
of dressed states of the form
$$\vert\psi\rangle\sum_{k=0}^{\infty}c_k(t)\vert 2k\rangle\eqno(10)$$
for even $n$ ; for odd $n$, $\vert 2k\rangle$ is replaced by 
$\vert 2k+1\rangle$. The possibility of emission/absorption of two
photons in the first order of interaction, i.e. linear in field intensity
becomes evident at once as a consequence of this model. Also, the commutation relations
(8) are gauge-independent, and hence from now on we can do away with the 
transversality restriction as mentioned after eqn.(3). 
\vskip 1pc
\centerline {\bf 3.\quad RESULTS AND DISCUSSION}
\vskip 8pt
Before we work out the possible consequences of this non-local field
in a typical atomic transition, let us note its formal similarity
with a {\it squeezed\/} state. We recall the definition
of the squeezed state [6] :
$$\vert\alpha,\zeta\rangle = D(\alpha)S(\zeta)\vert 0\rangle,\eqno(11)$$
where the {\it squeeze operator\/} $S(\zeta) = \exp [{1\over 2}\zeta^*a^2 
- {1\over 2}\zeta(a^{\dag})^2]$,
and the {\it displacement operator\/} $D(\alpha)=\exp 
{(\alpha a^{\dag}-\alpha^*a)} = e^{{-\vert\alpha\vert^2}
\over 2}e^{\alpha a^{\dag}}e^{-\alpha^*a}$. ($\alpha$ and $\zeta$ are arbitrary
complex numbers.) Putting $\alpha = 0$ gives the squeezed vacuum, 
while putting $\zeta = 0$ gives the coherent state. An expression 
for the squeezed vacuum in terms of the number states has been given by
Hollenhorst [7] ; putting $\zeta = re^{i\theta}$, this becomes
\vskip 1pc
\hfill $S(\zeta)\vert 0\rangle=(\cosh {r})^{1\over 2}\sum\limits_{n=0}^{\infty}
\left ({1\over 2} e^{i\theta}\tanh {r}\right 
)^n{\sqrt{2n!}\over {n!}}\thinspace\vert 2n\rangle$\hfill (12)
\vskip 1pc
Thus we see that the squeezed vacuum $S(\zeta)\vert 0\rangle$
has the same form as the expansion (10) above. The {\it two-mode squeezed
vacuum\/} [8], where the squeezing operator is given by the more generalised
form 
$$S(\zeta)=\exp[\zeta(\omega)a^{\dag}(\omega)a^{\dag}(2\Omega-\omega)
-\zeta^*(\omega)a(\omega)a(2\Omega-\omega)]$$
provides another example where the doubly occupied modes in (12) would
be replaced by singly occupied correlated mode pairs ; the signal and idler 
photons generated in a parametric down-conversion present themselves at once
as forming such a correlated pair.  We can therefore say that the
squeezing operator $S(\zeta)$ is mathematically identical (except for
coefficients/multiplying factors) to the evolution operator in a non-local 
field, and hence atoms interacting with a squeezed field and with an intense 
radiation field (for which a non-local description is valid) can be expected
to show the same non-classical features. We find below that it is indeed so.

We now work out two typical first-order transition matrix elements using this
non-local description for a radiation field. Some 
of these results have appeared before [2b], but are included here for the 
sake of completeness. First it must be mentioned that depending on the
photon flux and on the coherence properties of the laser field, one must
in general describe the field as 
$${\cal E} = a_L{\bf E}_L + a_N{\cal E}_N\eqno(13)$$
where the subscripts L and N refer to the local (usual) and the non-local
fields respectively ($a_L$ and $a_N$ are identical with $a_{\rm I}$ and 
$a_{\rm II}$ in eqn.(5)) ; the relative strength of the two coefficients 
would depend, essentially, upon the degree to which the response
of the atomic electron/s to the field appears phase-correlated.

We work within the dipole approximation, and formulate the typical matrix 
element of this two-photon absorption operator (I) for a single-electron 
atom, and (II) for a two-electron atom. 

(I) For simplicity we confine ourselves to a single radiation mode
$\vert n_{{\bf k}\lambda}\rangle$. Define
$$T_{\rm I}=\langle\psi_{l_fm_f}(x),n_{{\bf k}\lambda}-2\vert {\bf r}.{\cal
E}_N(x,x')\vert\psi_{l_i,m_i}(x'),n_{{\bf k}\lambda}\rangle\eqno(14)$$
where the $l$'s are the orbital angular momenta in the initial and final
states. We use equation (9b) for ${\cal E}_N(x,x')$ and note that the 
time integration
in (14) for the non-local interaction yields the required energy balance
$E_f=E_i+2\hbar\omega_{\bf k}$ {\it if and only if we use the retarded 
interaction\/} ; use of the advanced interaction fails to satisfy energy 
conservation. (See Appendix B for the mathematical details.) This remarkable 
result shows that causality as we understand it --- {\it ability of the past, 
but not of the future, to influence the present\/} --- follows as a
necessary condition for energy conservation in our non-local picture of the
electromagnetic field ; we have obtained a non-thermodynamic arrow of time
having a purely quantum nature.

Incidentally, energy conservation is satisfied even if we set the upper 
limit of the integration over $dt'$ (eqn.(B3), Appendix B) to
$t$ {\it exactly}, i.e. put $c=\infty$, which corresponds to an EPR-like
``quantum entanglement''. Thus we find, as had been mentioned in the 
beginning, that such situations need not be left out from our non-local
picture, and that causality in a quantum system need not be restricted by 
the special theory of relativity. However, if we go further in a closer 
analysis of the integration over $dt'$, we find that we can make an even 
stronger statement. Consider the case where $\vert{\bf r}-{\bf r}'\vert >ct$, 
in which case the upper limit in the $t'$-integral in (B3) becomes negative 
(remember that both ${\bf r}$ and ${\bf r}'$ extend over all space), 
and the integral vanishes, leading to non-conservation of energy !
Thus in our non-local picture of the quantum world, attempts to 
combine causality with the special theory of relativity are not only 
unnecessary, but can also lead to results which do not agree with 
experience ; one is immediately led to think of Bell's theorem. 
Note that such an EPR-like correlation between $x$ and $x'$ is also
indicated by the commutation relation (A3) in Appendix A, which gives
the necessary condition for {\it gauge invariance\/} of the non-local 
electromagnetic field tensor.

Finally we get (in atomic units)
$$T_{\rm I}= i{\sqrt{2\pi I}\over c}\langle\psi_f({\bf r})\vert {\bf r}.
\hat\epsilon(\hat r,\hat r')\vert\psi_i({\bf r}')\rangle\eqno(15)$$
As mentioned earlier, we use for $\hat\epsilon(\hat r,\hat r')$ a power series
in $\hat r$ and $\hat r'$ with undetermined coefficients ; the specific
form chosen is
$$\hat\epsilon(\hat r,\hat r')=(\hat r+\hat r')\sum_{n=0}^{\infty}
a_nP_n(\hat r.\hat r')\eqno(16)$$
(Symmetry demands that the coefficients of $\hat r$ and of $\hat r'$ be 
identical.) Putting $\psi_i({\bf r})=N_i{R_i(r)\over r}Y_{l_im_i}(\hat r)$ and
similarly for $\psi_f$ gives
$$T_{\rm I}= iN_iN_f\times {\sqrt{2\pi I}\over c}{\cal R}
({\cal I}_1+{\cal I}_2)\eqno(17)$$
where ${\cal R}$ is a radial integral, 
and the ${\cal I}$'s are two angular integrals, one of
which gives the selection rule $l_f=l_i$, while the other
gives $l_f=l_i$, $l_i\pm 2$. (See Appendix C for details.) 
Thus from (17) we obtain two-photon transitions of the type 
$S\to S$, $S\to D$\dots\thinspace linear in intensity.  We also note
that the squeezed vacuum corresponds to the limit $\alpha\to 0$ in eqn. (11) ;
an increase in $\alpha$, which results among other things in an increase  
in intensity, leads to a deviation of the squeezed field from the form
(12) to a situation where the summation over mode indexes runs over all
integers (instead of {\it only even or only odd integers\/} as in (10)),
so that in the first order, an increase in intensity in squeezed 
light-matter interaction causes, progressively, single-photon transitions 
instead of the two-photon transitions as was the case in the squeezed vacuum.
This result exactly matches the observation of Georgiades et al [1].
Other possible two-photon absorptions linear in intensity in a 
squeezed vacuum can be trivially predicted. Two possibilities immediately
come to mind that should be experimentally feasible : (i) two-photon 
ionisation of a Rydberg atom, and (ii) two-photon electron detachment 
from a negative ion, both with a squeezed vacuum. Indeed, a whole field of
non-destructive two-photon spectroscopy of atoms/molecules opens up. 

At this point we should remark that the formal similarity between a 
squeezed field and a laser field with strong second-order coherence
can become misleading when we consider the intensity variation of the
fields interacting with matter. As we shall soon see, the higher the
intensity, the more phase-correlated does the classical (thermal or laser) 
radiation field ``appear'' to two detectors (e.g. two atomic electrons), or 
one detector at two different times, rendering this non-local description 
the more valid. On the other hand, as we just saw, increase in intensity in a 
squeezed vacuum takes us back, so to say, into the classical domain (the 
two-photon transition becomes non-linear with increase in intensity).
 
\vskip 6pt
(II) For a two-electron atom we require a matrix element of the type
$$T_{\rm II}=\langle\psi_{l_fl'_f}(x,x'),n_{{\bf k}\lambda}-2\vert
({\bf r}+{\bf r}').{\cal E}_N(x,x')\vert\psi_{l_i,l'_i}
(x,x'),n_{{\bf k}\lambda}\rangle\eqno(18)$$
As before, the time integration in (18) yields the energy balance
$\delta(E_f,E_i+2\hbar\omega_{\bf k})$, leaving
$$T_{\rm II}=i{\sqrt{2\pi I}\over c}\langle\psi_f({\bf r},{\bf r}')\vert
({\bf r}+{\bf r}').\hat\epsilon(\hat r,\hat r')\vert\psi_i({\bf r},{\bf r}')
\rangle\eqno(19)$$

To evaluate $T_{\rm II}$ we use a correlated wavefunction of the form [9] 
$$\psi=R^{-{5\over 2}}\sum_{\mu}F_{\mu}(R)
\Phi_{\mu}(R;\alpha,\hat r,\hat r')\eqno(20)$$ 
where $R=\sqrt{r^2+r'^2}$ and $\alpha=\arctan {(r'/r)}$ are the
hyperspherical coordinates. 

Substituting eqns. (20) and (16) in (19) we obtain 
(see Appendix D for details) 
$$\eqalignno{T_{\rm II}&
= i{\sqrt{2\pi I}\over c}\int F_f^*(R)RF_i(R)\,dR \thinspace
\int A_fA_iP_{n_f}^{(a_f,b_f)}(\cos 2\alpha)(\cos\alpha)^{l_f+l_i}\cr 
&\times (\sin\alpha)^{l'_f+l'_i}(\cos\alpha+\sin\alpha)P_{n_i}^{(a_i,b_i)}
(\cos 2\alpha)\cos^2\alpha\sin^2\alpha\,d\alpha\cr
&\int\int\left [{\cal Y}^{M_f}_{L_fl_fl'_f}(\hat r,\hat r')
\right ]^*(1+P_1(\hat r.\hat r')\sum_{k=0}^{\infty}
a_kP_k(\hat r.\hat r'){\cal Y}^{M_i}_{L_il_il'_i}
(\hat r,\hat r')\,d\hat r\,d\hat r'&(21)\cr}$$
~~~~~We obtain the selection rules for $l_f$ and $l'_f$ from the two angular 
integrals, one involving $\sum_ka_kP_k(\hat r.\hat r')$ and the other 
involving $\sum_ka_kP_1(\hat r.\hat r')P_k(\hat r.\hat r')$. They are 
tedious but straightforward ; note that although the coefficients $a_k$ 
are unknown, the relative strength of the two terms can be analytically
obtained. For the first integral, both $\Delta l$ and $\Delta l'$ are even 
for even $k$ and odd for odd $k$, whereas for the second integral $\Delta l$ 
and $\Delta l'$ turn out to be odd for even $k$ and even for odd $k$.
Thus finally we see that parity is conserved, as in the single-electron
case.

For carrying out the integration over $d\alpha$ one can use the result [10]
$$\eqalignno{&\int^1_{-1}(1-x)^{\rho}(1+x)^{\sigma}P_n^{(\alpha,\beta)}(x)
P_m^{(\gamma,\delta)}(x)\,dx\cr
&= {2^{\rho+\sigma+1}(1+\alpha)_n(1+\gamma)_m\Gamma(\sigma+1)\Gamma(\rho+1)
\over{m!n!\Gamma(\rho+\sigma+2)}}
\sum_{r=0}^m{(-m)_r(1+\gamma+\delta+m)_r(1+\rho)_r\over
{r!(1+\gamma)_r(\rho+\sigma+2)_r}}\cr
&\times\thinspace _3F_2(-n,1+\alpha+\beta+n,1+\rho+r;1+\alpha,
2+\rho+\sigma+r;1)&(22)\cr}$$
On symmetry grounds, $\Delta n=n_f-n_i$ is always even ; sample calculations
show that the integral over $d\alpha$ is strongly peaked when $n_f=n_i$.
We see therefore that this two-photon transition {\it maintains the radial
correlation\/} in the two-electron atom.

We can now summarise the properties of the matrix element $T$ as follows :

(i) It is a two-photon excitation linear in the field strength ;

(ii) It conserves the parity of the atom ;

(iii) For a two-electron atom, it conserves the degree of radial correlation.

In view of eqn.(13) this double excitation will be running, in general, 
parallel to usual single-photon transitions in an experiment, and hence in
any multiphoton excitation process in a field having a strong second-order
coherence the overall transition would have the form $TTDDTD\dots$ or some
such random mixture ($D$'s are single-photon transition moments), tending
towards the extreme form $TTTT\dots$ as the field grows more intense.
If we start out with a closed-shell atom in an $S$-state, such a series
of $T$-transitions would result in the formation of strongly correlated
doubly excited states which are known to favour simultaneous double electron
escape (Wannier [11] ; see also the review articles of Fano [9] and Rau [12]). 
This is to be contrasted with {\it sequential\/} two-electron escape,
which does not involve any electron correlation.

We now demonstrate briefly that many of the features observed in
multiphoton double ionisation in atoms placed in strong laser fields can
be shown to follow as natural consequences of this non-local field model.
A fuller demonstration can be found in [2b]. As mentioned at the outset,
interaction of matter with a strong radiation field fits into this
non-local picture if two photons are absorbed by an atom within a time
interval $\delta t\ll{1\over\omega}$. For practical purposes this 
requirement can be replaced by a more convenient one, namely that in a
time ${1\over\omega}$, the number of photons flowing into and out of the
``volume'' V of the atomic shell occupied by the outer electron/s be
much larger than the number of these outer electrons. Thus for an 
$n$-(outer)-electron atom the condition reads
$${{VI}\over{c\omega^2}}\gg n\eqno(23)$$
On the other hand, when the correlated motion of two electrons is
involved (see discussion after eqn.(22)), we have to remember that 
the laser electric field should be significantly smaller than the Coulomb
fields within the atom, in order that the electrons can maintain their 
mutual correlation. Choosing a field of order $10^{-2}$ a.u. gives us a
crude estimate for an upper limit
$$I_{max}\sim 10^{13}~{\rm W/cm^2}\eqno(24)$$
to the laser intensity upto which we can expect electron correlation in
a two-electron atom to be maintained.

We now examine some physical consequences of our model QED and compare
them with observed features in Multiphoton Ionisation (MPI) and
Above-threshold Ionisation (ATI) experiments.  Earlier, the occurrence
of a $TT..$-like chain was shown to lead to the possibility of {\it direct 
double ionisation\/} (as opposed to {\it sequential\/} double ionisation) 
[13] ; we now see that the
possibility can materialise if and only if both the conditions (23) and (24)
are fulfilled. Simple considerations show that for short wavelengths the
two conditions can become mutually exclusive, thus ruling out the possibility
of Direct Double Ionization (DDI), whereas for long wavelengths both 
conditions (23) and (24) can be satisfied over a broad range of intensities. 
There is a caveat, though ; for {\it very large\/} wavelengths the condition 
(23) can be satisfied at quite low intensities, but the rather large number 
of $D$'s in the multiphoton excitation chain $TTDDTD\dots$ 
will tend to spoil the electron correlation
pattern. Also, for a given power output, the peak intensity in short
laser pulses is higher than in longer ones, and hence we should expect
DDI to occur more readily with shorter pulses than with long pulses. 
This accords with the observations of L'Huillier et al [14] who measured
multiphoton ionisation in Xenon using both nanosecond and picosecond 
pulsed lasers at 1064 nm and 532 nm, and those of Agostini and Petite [15]
and Delone et al [16] who measured MPI of Ca at 1064 nm with picosecond
pulses and nanosecond pulses respectively. Furthermore, Yergeau et al [17]
found that in MPI of rare gases using a CO$_2$ laser ($\lambda=9.55\mu$
and $10.55\mu$), the multiply charged ions are formed {\it sequentially}.

In a strong laser field,
another interesting effect stems from the photon number-phase uncertainty
relation $\Delta N\Delta\phi\ge 1$. The atomic electron/s during the 
two-photon excitation is/are acting as a device that measures the 
incident radiation phase to within an accuracy $\Delta\phi=\omega\delta t$.
Consequently the photon number would be uncertain by $\Delta N\sim {1\over
{\omega\delta t}}$, and this would shift the photoelectron spectrum by an
amount $\Delta N\hbar\omega$, resulting in the suppression of $\Delta N$
lower-order peaks. This is a very well-known phenomenon in ATI (e.g. [18]) ; 
attempts to explain it have so far met with only partial success. Note that 
in the case of two electrons, they can act individually as such phase-measuring
devices and no correlation is required ; as such, this feature is not
restricted by the condition (24). Indeed, peak suppression (also known as
quenching) has been observed at very high intensities ($\sim 10^{16}$ 
W/cm$^2$).

Use of circularly polarised light lends a particularly interesting twist
to this aspect. A circularly polarised photon can be regarded as a 
superposition of two linearly polarised components $90^o$ apart in phase,
and hence even if two circularly polarised photons arrive with a phase 
gap $\omega\delta t={\pi\over 2}$ between them, the $x$-component of one 
photon and the $y$-component of the other can still be in exact phase 
correlation. This implies that the coherence condition $\delta t\ll {1\over
\omega}$ can be relaxed to $\delta t\sim {1\over\omega}$, i.e. it would
be satisfied at much smaller intensities. Thus at any given intensity,
many more lower-order ATI peaks can be expected to be quenched by
circularly polarised light than by linearly polarised light. This feature
has indeed been observed (e.g. Bucksbaum et al [19], Bashkansky et al [20]).
Interestingly, this also allows us to {\it predict\/} that with a circularly 
polarised laser, onset of DDI should occur in MPI experiments at lower 
intensities than with linearly polarised lasers.

In conclusion, we find that our non-local model of QED gives us a 
correlated two-photon absorption which, unlike the conventional four-field
correlation function, occurs in the first order of perturbation and is
therefore linear in intensity. A single first-order matrix element [$T_{\rm 
I}$, eqn.(17)] exactly reproduces the experimental finding of Georgiades et 
al [1], while in higher orders we obtain results that match very well with
several experimental findings in MPI and ATI of atoms. From a detailed
work-out of the non-local QED we obtain a non-thermodynamic arrow of time,
and we also find that correlations in a quantum system must necessarily be 
of the EPR-type, a result that agrees with Bell's theorem.
This non-local model QED can certainly be further improved upon by way of 
rigour and sophistication, but we believe that, basically, we have been 
able to demonstrate its usefulness. 
\vfill\eject
\centerline {\bf APPENDIX A}
\vskip 1pc
In this appendix we use relativistic notation, with units $\hbar=c=1$.
Let $A^0$ and {\bf A} be the scalar and vector potentials of an
electromagnetic field. In field theory, the four-potential $A^{\mu}
\equiv(A^0,{\bf A})$ is usually introduced through imposition of the 
requirement of covariance under a $U(1)$
gauge transformation (of the type $e^{-ie\Lambda(x)}$), 
and this is done via construction of the covariant derivative
$D_{\mu}=\partial_{\mu}+ieA_{\mu}$ ($A_{\mu}=g_{\mu\nu}A^{\nu}$,
$g_{\mu\nu}$ being the metric tensor) and replacement of the partial 
derivatives in the Lagrangian (e.g.\thinspace of 
a complex scalar field $\phi$) by these covariant derivatives. From this
gauge-covariant Lagrangian $D_{\mu}\phi D^{\mu}\phi$ we obtain the
{\it interaction Lagrangian\/} $-ej^{\mu}A_{\mu}$, where $j^{\mu}=i(\phi^*
\partial^{\mu}\phi - \phi\partial^{\mu}\phi^*)$. (Of course, these $A_{\mu}$'s
need not be a dynamical field but can be a pure gauge field as well [21] ;
however, this does not concern us here.)

A significant point that is often lost sight of is that, in order to 
achieve gauge covariance, {\it it is not a necessary condition\/}
that this interaction Lagrangian be {\it local} ; indeed,
we can write down a non-local Lagrangian
in a similar manner. Consider a non-local $U(1)$ gauge transformation 
of the kind $\exp (-ie\Lambda\vert x\rangle\langle x'\vert )$. It is easily
seen that a covariant derivative can be constructed as $D_{\mu}=\partial_{\mu}
+d_{\mu}+ie{\cal A}_{\mu}\vert x\rangle\langle x'\vert $, (where $d_{\mu}
\equiv \partial/\partial x'^{\mu}$ and we have written ${\cal A}_{\mu}\vert 
x\rangle\langle x'\vert\equiv g_{\mu\nu}{\cal A}^{\nu}\vert x
\rangle\langle x'\vert$), provided that 
${\cal A}_{\mu}$ transforms under the gauge transformation as
$${\cal A}_{\mu} \to {\cal A}_{\mu}+\partial_{\mu}\Lambda\vert x\rangle\langle
x'\vert + d_{\mu}\Lambda\vert x\rangle\langle x'\vert .\eqno({\rm A}1)$$
It can be readily verified that 
this transformation property is consistent with the requirement of
{\it gauge invariance\/} of the non-local (electromagnetic) field tensor
$${\cal F}^{\mu\nu}\vert x\rangle\langle x'\vert=(\partial^{\mu}+d^{\mu}) 
{\cal A}^{\nu}\vert x\rangle\langle x'\vert - (\partial^{\nu}+d^{\nu})
{\cal A}^{\mu}\vert x\rangle\langle x'\vert \eqno({\rm A}2)$$
under this new gauge transformation, {\it provided that}
$$\left [\partial^{\mu},d^{\nu}\right ]=0.\eqno({\rm A}3)$$
~~~~Ultimately, we have to replace $A_{\mu}$ 
in the above-defined interaction Lagrangian by 
$g_{\mu\nu}{\cal A}^{\nu}\vert x\rangle\langle x'\vert$. 
We see, therefore, that our non-local potential {\it 
is consistent with the basic requirement of gauge covariance 
of the Lagrangian\/}. Our eqn.\thinspace (1) for the matrix element 
$M_{fi}(t)$ follows directly from this non-local interaction.

The nature of the correlation between $x$ and $x'$ in this formulation
is laid down by the above commutation relation (A3), which indicates
that $x$ and $x'$ must be completely independent, i.e.\thinspace there is
no restriction that their interval be time-like. We postpone a detailed
discussion of this point till later ; in connection with eqn.\thinspace
(14) we shall show (see second paragraph after the eqn.\thinspace
in text) that energy conservation at the atomic level requires that 
this correlation must, of necessity, be of the EPR-type.
\vfill\eject
\centerline {\bf APPENDIX B}
\vskip 1pc
We show in this appendix that only the retarded interaction can lead to 
energy conservation. The time-integral in (14), with retarded correlation,
is given by
$$\int_{-\infty}^{\infty}e^{{i\over\hbar}(E_ft-\hbar\omega t)}\,dt
\int_{-\infty}^{t-\rho/c}e^{-{i\over\hbar}(E_it'+\hbar\omega t')}\,dt'
=\int_{-\infty}^{\infty}f(t)\,dt\int_{-\infty}^{t-\rho/c}
g(t')\,dt'\quad ({\rm say})\eqno({\rm B}1) $$
($\rho=\vert{\bf r}-{\bf r}'\vert $), while with the advanced correlation 
we have
$$\int_{-\infty}^{\infty}f(t)\,dt\int_{t+\rho/c}^{\infty}
g(t')\,dt'\eqno({\rm B}2) $$
The expression (B1) can be written as a sum of two integrals :
$$\int_{-\infty}^0f(t)\,dt\int_{-\infty}^{t-\rho/c}g(t')\,dt'
+\int_0^{\infty}f(t)\,dt\int_{-\infty}^{t-\rho/c}g(t')\,dt'\eqno({\rm B}1')$$
while the expression (B2) can be written as
$$\int_{-\infty}^0f(t)\,dt\int_{t+\rho/c}^{\infty}g(t')\,dt'
+\int_0^{\infty}f(t)\,dt\int_{t+\rho/c}^{\infty}g(t')\,dt'\eqno({\rm B}2')$$
Elementary considerations show that the first integral in (B$1'$) and the
second integral in (B$2'$) contribute nothing, 
while the other two integrals are (to within
factors) given by $\delta(E_f,E_i+2\hbar\omega)$. Energy conservation in
the two-photon transition thus results, as it were, when we sum up our 
``samplings'' of the phase of the matrix element over extended periods
of time. 

We now come to the question of the significance of the {\it limits\/} 
of the integrals over $dt$. The matrix element is, so to say, {\bf created} 
when we have photoexcited the atom ; let us agree to define that this act 
has been done at the instant $t=0$. The crucial point to remember is that
although the photoexcitation is not an irreversible process (the atom can
certainly be de-excited again), {\it our acts of exciting and subsequently
observing the atom are irreversible in time\/}. Therefore
we can sample the phase (or whatever else we like to) of the
excitation matrix element only at times $t\geq 0$. Thus the first integral in
(B$2'$), although non-vanishing, is actually non-physical (it is just as
meaningful as sampling the behavioural pattern of a baby before the
baby is born), and we are left with only one non-vanishing integral in (B$1'$)
$$\int_0^{\infty}e^{{i\over\hbar}(E_ft-\hbar\omega t)}\,dt
\int_{-\infty}^{t-\rho/c}e^{-{i\over\hbar}(E_it'+\hbar\omega t')}\,dt'
\eqno({\rm B}3)$$
Thus energy conservation will be satisfied if and only if temporal 
correlation is provided by the past, and not by the
future ; we have derived an arrow of time in quantum physics as a necessary
condition for energy conservation. 
\vfill\eject
\centerline {\bf APPENDIX C}
\vskip 1pc
Elementary analysis shows that
$${\cal R}=\int_0^{\infty}R_i(r)r^2\,dr\int_0^{\infty}R_f(r)r\,dr,$$
$${\cal I}_1=\sum_{n=0}^{\infty}{4\pi a_n\over {2n+1}}
\sum_{\nu=-n}^n\int Y_{l_fm_f}^*
(\hat r)Y_{n\nu}(\hat r)\,d\hat r\thinspace\int Y_{n\nu}^*(\hat r')Y_{l_im_i}
(\hat r')\,d\hat r',$$
$$\eqalignno{{\rm and}\quad{\cal I}_2&=\sum_{n=0}^{\infty}
{16\pi^2a_n\over {3(2n+1)}}
\sum_{\nu=-n}^n\thinspace\sum_{m=-1}^1\int Y_{l_fm_f}^*(\hat r)Y_{1m}(\hat r)
Y_{n\nu}(\hat r)\,d\hat r\cr
&\times\int Y_{1m}^*(\hat r')Y_{n\nu}^*(\hat r')
Y_{l_im_i}(\hat r')\,d\hat r'\cr}$$
The first angular integral gives $l_f=l_i$, while the second integral, which
can be evaluated using the Gaunt formula, vanishes unless $n=l_f\pm 1$ and
$n=l_i\pm 1$. Combining these two we get the selection rules $l_f=l_i$, 
$l_i\pm 2$.
\vfill\eject
\centerline {\bf APPENDIX D}
\vskip 1pc
The materials in this appendix are standard, and are included only for the
sake of completeness. More details may be found in [9] (we have introduced
some minor changes in notation). The Schr\"odinger equation for a two-electron
atom in hyperspherical coordinates takes the form
$$\left [{\hbar^2\over 2m}\left (-{\partial^2\over \partial R^2}+
{\Lambda^2\over R^2}\right )+V-E\right ]\Psi(R,\alpha,\hat r,\hat r') = 0
\eqno({\rm D}1)$$
where the ``grand angular momentum operator''
$$\Lambda^2 = -{\partial^2\over {\partial\alpha^2}}-{1\over 4}+
{l(l+1)\over{\cos^2\alpha}}+{l'(l'+1)\over{\sin^2\alpha}}\eqno({\rm D}2)$$
Also, $V = -{Z\over{r}}-{Z\over{r'}}+{1\over{\vert{\bf r}-{\bf r}'
\vert}}$, the total Coulomb interaction within the atom, and $m$ is the
electron mass. 
The channel eigenfunctions $\Phi$'s in (20) 
are obtained by configuration mixing in a basis of eigenfunctions of the 
grand angular momentum operator (D2).
The eigenfunctions of this operator are
$$\phi_{nll'LM} = A(\cos\alpha)^{l}(\sin\alpha)^{l'}
P_n^{(a,b)}(\cos 2\alpha){\cal Y}^M_{Lll'}(\hat r,\hat r')
\eqno({\rm D}3)$$
where $A$ is a normalisation constant, $a = l'+{1\over 2}$, 
$b = l+{1\over 2}$, 
$${\cal Y}^M_{Lll'} = \sum_{m=-l}^l\sum_{m'=-l'}^{l'}(ll'mm'\vert ll'LM)
Y_{lm}(\hat r)Y_{l'm'}(\hat r'),$$
\hfill $P_n^{(a,b)}(x)={(a+1)_n\over{n!}}\thinspace _2F_1\left (-n,a+b+n+1;a+1;
{{1-x}\over 2}\right )$\hfill (D4)
\vskip 6pt
\noindent is the Jacobi polynomial, and $n$ is the radial 
correlation quantum number, the same as $n_{rc}$ of Fano [9]. The eigenvalues 
of $\Lambda^2$ are given by $\lambda(\lambda+1)\hbar^2$, where
$\lambda = 2n+l+l'+{3\over 2}$.
Thus for each channel $\mu$ in (20) we can write
$$\Phi_{\mu}=\sum_j C_{\mu j}\phi_j(\alpha,\hat r,\hat r')$$
where the running index $j\equiv (n,l,l',L,M)$ collectively. 

In hyperspherical coordinates the total Coulomb potential $V$ is given by
$$V = {e^2\over R}C(\alpha,\hat r,\hat r'),\quad
C = -{Z\over{\cos\alpha}}-{Z\over{\sin\alpha}}+{1\over {(1-\hat r.\hat r'
\sin 2\alpha)^{1/2}}}\eqno({\rm D}5)$$
The hyperradial functions $F_{\mu}(R)$ in (20) are given by the
coupled equations
$$\left [{\hbar^2\over 2m}\left (-{d^2\over {dR^2}}+{\lambda(\lambda+1)\over
{R^2}}\right )-E\right ]F_{\mu}(R) + {e^2\over R}\sum_{\mu'}V_{\mu\mu'}
F_{\mu'}(R)\delta_{LL'}\delta_{MM'} = 0\eqno({\rm D}6)$$
Restricting ourselves to a single configuration in (20) and substituting
eqn.(16) in (19), we obtain eqn. (21) for $T_{\rm II}$.

\vfill\eject
\noindent {\bf REFERENCES :}
\vskip 6pt
\noindent [1] N P Georgiades, E S Polzik, K Edamatsu, H J Kimble and
A S Parkins, Phys. Rev. Lett. {\bf 75}, 3426 (1995). See also Z Ficek and 
P D Drummond, Phys. Today, Sept. 1997, p.34\par
\noindent [2](a) T K Rai Dastidar and K Rai Dastidar, XI Int.Conf.At.Phys. 
(Paris, 1988) p. I-23 ;\par
\noindent (b) K Rai Dastidar in {\it Advances in Atomic and  Molecular 
Physics\/}, ed. M S Z Chaghtai (Today's
and Tomorrow's Publishers, New Delhi, India 1992) p. 49. Some modifications 
over these early papers have been carried out in the present paper in the 
detailed work-out of the non-local ansatz ; however, the conclusions 
in these earlier works are not affected.\par
\noindent [3] J Gea-Banacloche, Phys.Rev.Lett. {\bf 62}, 1603 (1989)\par
\noindent [4] J Javanainen and P L Gould, Phys. Rev A {\bf 41}, 5088
(1990)\par
\noindent [5] I M Gelfand and G E Shilov, {\it Generalised Functions\/}
Vol. 1, Chap. 2, eq. 1.3(1) (Academic Press, 1964)\par
\noindent [6] See, e.g. C M Caves, Phys.Rev.D {\bf 23}, 1693 (1981)\par
\noindent [7] J N Hollenhorst, Phys.Rev.D {\bf 19}, 1669 (1979)\par
\noindent [8] V Buzek, P L Knight and I K Kudryavtsev, Phys.Rev.A {\bf 44},
1931 (1991) and references therein.\par
\noindent [9] U Fano, Rep.Prog.Phys. {\bf 46}, 97 (1983)\par
\noindent [10] R C Varma and J S Bharadwaj, Proc.Natl.Acad.Sci.India {\bf 54A},
422 (1984)\par
\noindent [11] G Wannier, Phys.Rev. {\bf 90}, 817 (1953)\par
\noindent [12] A R P Rau, Phys.Rep. {\bf 110}, 369 (1984)\par
\noindent [13] In an MPI experiment where ion counts are plotted against
laser intensity, this feature is characterised 
by the appearance of a doubly charged ion signal {\it before\/} the 
saturation of the singly charged ion signal. See e.g. K Rai Dastidar in {\it 
Atomic and Molecular Physics\/}, eds. D K Rai and D N Tripathi (World 
Scientific, Singapore 1987), p. 323.\par
\noindent [14] A L'Huillier, L A Lompre, G Mainfray and C Manus, J.Phys.B 
{\bf 16}, 1363 (1983) ; Phys. Rev. A {\bf 27}, 2503 (1983)\par
\noindent [15] P Agostini and G Petite, J.Phys.B {\bf 17}, L811 (1984)\par
\noindent [16] N B Delone, V V Suran and B A Zon, in {\it Multiphoton
Ionization of Atoms\/} eds. S L Chin and P Lambopoulos (Acad. Press, New York
1984) p. 247\par
\noindent [17] F Yergeau, S L Chin and P Lavigne, J.Phys.B {\bf 20}, 723
(1987)\par
\noindent [18] M Crance, Phys.Rep. {\bf 144}, 117 (1987)\par
\noindent [19] P H Bucksbaum, M Bashkansky, R P Freeman, T J McIlrath and
L F DiMauro, Phys. Rev. Lett. {\bf 56}, 2590 (1986)\par
\noindent [20] M Bashkansky, P H Bucksbaum and D Schumacher, Phys.Rev.Lett.
{\bf 59}, 274 (1987)\par
\noindent [21] T K Rai Dastidar and K Rai Dastidar, Mod.Phys.Letts. 
{\bf A10}, 1843 (1995)
\bye